\documentclass[eqsecnum,aip,jmp, amsmath,amssymb,reprint]{revtex4-1}
\usepackage[T1]{fontenc}
\usepackage{newpxtext,newpxmath,mathrsfs}
\usepackage[sans]{dsfont}
\usepackage{tikz}
\usetikzlibrary{cd}
\usepackage{hyperref}
\def\inner{\,{\vcenter{
      \hbox{ \kern 3pt
        \vrule width 0.43pt height 5pt}
      \hrule height 0.43pt}}\,}
\begin{document}
\baselineskip 12.5pt
\begin{titlepage}
\title{Quantization of Pseudoclassical Systems in the Schr\"odinger Realization}
         
\author{Theodore J. Allen}
\email{tjallen@hws.edu}
\author{Donald Spector}
\email{spector@hws.edu}
\affiliation{Department of Physics, Hobart and William Smith Colleges \\
Geneva, New York 14456 USA }
\author{Christopher Wilson}
\email{cww78@cornell.edu}
\affiliation{Department of Physics, Cornell University,
Ithaca, New York, 14853 USA}

\date{\today}
\thispagestyle{empty}
\begin{abstract}
 We examine the quantization of pseudoclassical dynamical systems, models
 that have classically anticommuting variables, in the Schr\"odinger
 picture. We quantize these systems, which can be viewed as classical
 models of particle spin, using the generalized Gupta-Bleuler method as well as
 the reduced phase space method in even dimensions. With
 minimal modifications, the standard constructions of Schr\"odinger
 quantum mechanics of constrained systems work for pseudoclassical
 systems. We generalize the standard Schr\"odinger 
 norm and implement the correct adjointness
 properties of observables and constraints. We construct the state space corresponding to spinors as
 physical wave functions of anticommuting variables, finding that there are superselection sectors in
 both the physical and ghost subspaces. The physical states are isomorphic to those of 
 the Dirac-K\"ahler formulation of fermions though the inner product in Dirac-K\"ahler theory is not equivalent to ours. 
\end{abstract}
\maketitle

\end{titlepage}
\section{Introduction}
Anticommuting variables, also called Grassmann numbers, have a long history
in theoretical physics,
\cite{Schwinger:1953qd,Martin:1959fp,Klauder:1960kt,Berezin:1961ct,Berezin:1966nc}
with applications ranging from the path integral formulation of fermions to
superspace constructions for supersymmetric theories. Pseudoclassical
mechanics, which incorporates anticommuting dynamical variables, arises as
the $\hbar \to 0$ classical limit\cite{Casalbuoni:1976tz,Berezin:1976eg} of quantum mechanical systems with spin.
{Despite the key role of
anticommuting variables in theoretical physics, the Schr\"odinger picture
approach to such systems heretofore has not received full attention. We aim to remedy
this oversight.}

In their renowned paper on the use of anticommuting variables to describe
relativistic and non-relativistic spin degrees of freedom, Berezin and
Marinov\cite{Berezin:1976eg} posit a three dimensional real vector-valued
anticommuting variable $\xi_k$ with the real Grassmann-even action
\begin{equation}\label{eq:BerezinAction} 
  S = \int dt\, \left[\frac{1}{2} \tilde \omega_{kl} \xi_k \dot \xi_l - H(\xi)\right] \> ,
\end{equation}
with $H$ a real Grassmann-even function and $\tilde \omega$ an imaginary symmetric $3\times3$ matrix, to describe
the non-relativistic spin degrees of freedom of a spin-1/2
particle. The matrix $\tilde \omega$ can be reduced to 
\begin{equation}\label{eq:SimpleForm}
\tilde\omega_{kl} = i\delta_{kl} 
\end{equation}
by a linear transformation of variables. With this choice
the kinetic term of the action is $O(3)$ invariant, and the full action will be
likewise if the function $H(\xi)$ is $O(3)$ invariant. 

Berezin and Marinov note from the form of the action that the
variables $\xi_k$ may be taken as  phase-space coordinates and then define a
Poisson bracket from the {(ortho)}symplectic form $\tilde \omega_{kl}$ that gives the 
correct equations of motion. After quantization, the operators $\hat \xi_k$ 
corresponding to the pseudoclassical variables $\xi_k$ become the generators 
of the Clifford algebra with three Euclidean generators and satisfy the Pauli matrix 
anticommutation relations. Consistent with their abstract approach to 
mechanics, Berezin and Marinov appeal to the representation theory of 
Clifford algebras, and take the space of states to be the essentially unique 
irreducible representation of that algebra, which is the space of two-component
spinors.

{While this approach is certainly elegant and efficient, as a phase space method
it bypasses the quantization on configuration space and 
provides no insight into the issues of dealing with first-order actions}
on configuration space. When particle position is considered as an additional 
configuration space variable, as Berezin and Marinov\cite{Berezin:1976eg} also do in their paper,
{the resulting actions have not only a global rotational (or Lorentz) invariance that acts on both particle position and the Grassmann-odd coordinates, but also 
a world-line supersymmetry relating the Grassmann-even and -odd coordinates, strongly suggesting their treatment on an equal footing.}

Our purpose in this paper is to analyze pseudoclassical systems in $D$ dimensions 
of the form introduced by Berezin and Marinov through the explicit application of 
Dirac's methods for constrained systems and the use of the Schr\"odinger 
representation for the quantum states and their norms, assuming that all the  
$\xi_i$ variables are configuration space coordinates. Certainly, the approach of taking the 
configuration space to be the full set of anticommuting variables is not new, having been
explored, for example, by Manko\v{c} Bor\v{s}tnik\cite{MankocBorstnik:1992np,MankocBorstnik:1993ia} and Manko\v{c} Bor\v{s}tnik and Nielsen,\cite{MankocBorstnik:1999th} {but our approach to implementing quantization in the full Schr\"odinger representation does appear to
be new.}
{Our work offers} concrete insights 
into both pseudoclassical mechanics and constrained quantization.  We make as close
an analogy with the standard Schr\"odinger picture as we can by using the calculus 
of anticommuting variables \cite{Faddeev:1980be,Berezin:1966nc,DeWitt:1992cy} in the standard constructions.

We denote left and right derivatives with respect to an anticommuting variable $\xi$ as 
$\partial^L/\partial\xi$ and $\partial^R/\partial\xi$ respectively.

\section{$D=2$ anticommuting variables}\label{sec:evendims}

%\subsection{Two-variable Pseudoclassical System}

\subsection{Pseudoclassical Hamiltonian System}
As a system with one anticommuting coordinate is something of a special case, {which we will
address later,} we start by analyzing in detail
the system with two anticommuting coordinates, proceed to the general case of an even number of anticommuting coordinates, 
and then come back to the one, three, and general odd-dimensional cases later.

Barring for the moment actions with non-dynamical Grassmann-odd parameters, the simplest non-trivial 
pseudoclassical action for two real Grassmann-odd coordinates 
$\xi_1$ and $\xi_2$  has the essentially unique Lagrangian,
\begin{equation}\label{eq:Lagrangian}
L=\frac{i}{2}(\xi_1 \dot\xi_1 + \xi_2 \dot\xi_2) + i \omega \xi_1\xi_2 \>  ,
\end{equation}
which is also the most general action invariant under the rotation group $SO(2)$. Here $\omega$ is
a real Grassmann-even constant. (In the form of (\ref{eq:BerezinAction}),  $H(\xi) 
= -\frac{i}{2}\epsilon_{ij}\xi_i\xi_j$ is rotationally invariant and unique up to the 
addition of a real Grassmann-even constant.)  The Euler-Lagrange equations of 
motion that follow from (\ref{eq:Lagrangian}),
\begin{equation}\label{eq:ELone}
\frac{d}{dt} \left({\partial^R L\over \partial \dot\xi_i}\right) =
{\partial^R L\over \partial \xi_i} \> ,
\end{equation}
are
\begin{equation}\label{eq:EoMone}
\dot\xi_i = \omega (\xi_1\delta_{i2} - \xi_2\delta_{i1}) = - \omega \epsilon_{ij}\xi_j \> .
\end{equation}

The Hamiltonian description, with Poisson brackets
\begin{equation}\label{eq:PBs}
\left\{f,g\right\} = \sum_{i=1,2} \left( {\partial^R f\over \partial
  \xi_i}{\partial^L g \over \partial\pi_i} + {\partial^R f \over
  \partial\pi_i}{\partial^L g \over \partial\xi_i} \right)  
\end{equation}
 and canonical momenta,
\begin{equation}\label{eq:canonicalmomenta}
\pi_i = {\partial^R L \over \partial \dot\xi_i} = \frac{i}{2}\xi_i 
\end{equation}
that do not depend on velocities $\dot\xi_i$, is complicated by the 
presence of constraint functions on the phase-space
\begin{equation}\label{eq:constraintsone}
\varphi_i = \pi_i - \frac{i}{2} \xi_i \approx 0 \> ,
\end{equation}
which are second-class in Dirac's\cite{Dirac:1964:LQM} 
classification, because they do not have vanishing Poisson 
brackets with themselves. These constraints reduce the dimension of the
phase space from four to two. Dirac's methods for analyzing both classical and quantum
constrained systems are well explained in the literature.\cite{Dirac:1964:LQM, Hanson:1976cn, Sundermeyer:1982gv, Gitman:1990qh, Govaerts:1991gd, Henneaux:1992ig}

On the physical phase space defined by the constraints (\ref{eq:constraintsone}), 
the Hamiltonian is equal to
\begin{equation}\label{eq:Hzero}
H = \pi_i \dot\xi_i - L = - i\omega\xi_1\xi_2 \> , 
\end{equation}
but the evolution of the system on the physical phase space defined by the
constraints (\ref{eq:constraintsone}) must stay in that phase space. In other words,
the constraints must be conserved in time. The most general Hamiltonian that
agrees with (\ref{eq:Hzero}) on the physical phase space is
\begin{equation}\label{eq:class}
H^\prime = - i\omega\xi_1\xi_2 + \lambda_i \varphi_i \> ,
\end{equation}
where the $\lambda_i$ are Grassmann-odd phase space functions.

%The equations of motion arising from the Hamiltonian (\ref{eq:class}) are
%\begin{eqnarray}\label{eq:CanonicalEoMone}
%\dot \pi_i = \left\{\pi_i,H\right\} &=& \frac{i}{2} \lambda_i - i\omega
%\epsilon_{ij}\xi_j + {\partial^L \lambda_j\over
 % \partial\xi_i}\varphi_j \approx  \frac{i}{2} \lambda_i - i\omega
%\epsilon_{ij}\xi_j  , \cr
%\dot \xi_i = \left\{\xi_i,H\right\} &=& - \lambda_i + {\partial^L
%  \lambda_j\over \partial \pi_i} \varphi_j \approx -\lambda_i .
%\end{eqnarray}
The coefficients $\lambda_i$ are determined by requiring that the
constraints remain zero on the reduced phase space,
\begin{equation}\label{eq:lambdas}
\dot\varphi_i = \left\{\varphi_i,H^\prime\right\} = i(\lambda_i -
\omega \epsilon_{ij} \xi_j ) + \left({\partial^L \lambda_j\over
  \partial\xi_i} - \frac{i}{2}{\partial^L
  \lambda_j\over \partial \pi_i}\right) \varphi_j \approx 0 \> .
\end{equation}
Equation~(\ref{eq:lambdas}) can be made to hold identically, not just weakly, when
we choose
\begin{equation}
  \lambda_i = \frac{3}{4} \omega \epsilon_{ij}\xi_j -
  \frac{i}{2}\omega\epsilon_{ij}\pi_j  \> ,
\end{equation}
from which we have
\begin{eqnarray}\label{eq:Hthree}
  {H}^\prime & = & -\frac{i}{4}\omega \xi_1\xi_2 + \frac{1}{2} \omega
  (\xi_2\pi_1 - \xi_1\pi_2) + i\omega \pi_1 \pi_2 \cr
    & = & i\omega\left(\pi_1 + \frac{i}{2}\xi_1\right)\left(\pi_2 + \frac{i}{2}\xi_2\right) \>  .
\end{eqnarray}

%It is enough for consistency to fix $\lambda_i =
%\omega\epsilon_{ij}\xi_j$, 
%yielding the physical Hamiltonian
%\begin{equation}\label{eq:Htwo}
%H_{\rm phys} = \omega\left(\xi_2\pi_1 - \xi_1\pi_2\right) =
%-\omega\epsilon_{ij} \xi_i \pi_j\, ,
%\end{equation}
%which leads to the same equations of
%motion as the Euler-Lagrange equations of Eq.~(\ref{eq:EoMone}), at least
%on the constraint surface:
%\begin{equation}\label{eq:EoMtwo} 
%\dot\xi_i = \left\{\xi_i,H_{\rm phys}\right\} = {\partial^L H_{\rm phys}
  %\over \partial \pi_i}\approx - \omega \epsilon_{ij}\xi_j. 
%\end{equation}

\subsection{Wave functions,  involution, and operators}

Generalizing from the standard quantum mechanics, we take the wave functions for 
the quantum states to be complex Grassmann-valued functions of the coordinates $\xi_i$,
defined through their power series,
\begin{equation}\label{eq:StateSeries}
\psi = \psi(\xi_1,\xi_2) = \psi_0 + \psi_1\xi_1 + \psi_2 \xi_2 + \psi_3 \xi_1\xi_2 \> ,
\end{equation}
with complex-valued (Grassmann-even) coefficients $\psi_i$.

While the Grassmann coordinates $\xi_i$ are taken to be real, $\xi_i^*=\xi_i$, because of the properties 
of the involution
\begin{eqnarray}\label{eq:starinvolution}
(\xi^{*})^* &=& \xi \> , \cr
  (\xi\theta)^*& =& \theta^* \xi^* 
\end{eqnarray}
that generalizes complex conjugation to Grassmann variables,  the complex conjugate of a product 
of two odd elements of a Grassmann algebra is $ (\xi\theta)^* = \theta^* \xi^* = - \xi^*\theta^*$.
This involution property, when combined with anticommutativity, yields the unfamiliar 
result that the product of two real anticommuting numbers is imaginary, as is the product of two imaginary anticommuting numbers.  The properties of
classical variables under complex conjugation carry over to the
adjointness properties of their corresponding quantum operators.

In the Schr\"odinger representation the phase space variables are replaced by the operators
\begin{eqnarray}\label{eq:Opsone}
\hat{\xi}_i &=& \xi_i  \> ,\cr
\hat{\pi}_i &=& i\hbar {\partial^L\over\partial\xi_i}  \> .
\end{eqnarray}
In what follows, we use units where $\hbar = 1$.

\subsection{Inner product and physical states}

In keeping with the analogy to the commuting case, we take the inner product 
to be proportional to the $SO(2)$ invariant Grassmann integral over 
configuration space. In the notation of (\ref{eq:StateSeries}),
\begin{equation}\label{eq:NaiveInner}
\int \phi^* \psi\, d\xi_1 d\xi_2 = \phi_3^*
\psi_0 + \phi_2^* \psi_1 - \phi_1^* \psi_2 - \phi_0^* \psi_3 \>  .
\end{equation}
With an explicit factor of $i$, the Grassmann integral (\ref{eq:NaiveInner}) gives an inner product,
\begin{equation}\label{eq:SchrodNorm}
  \langle \phi|\psi \rangle = i \int \phi^* \psi\, d\xi_1 d\xi_2 = \langle \psi | \phi \rangle^* \> ,
\end{equation}
that yields a manifestly real norm, but one that is not positive definite on the
full space of wave functions (\ref{eq:StateSeries}).  However, as in the 
case of gauge theories, the inner product need only be positive definite on 
the space of physical states. States of non-positive norm are 
unphysical ``ghost'' states.

\subsection{Generalized Gupta-Bleuler Quantization}

In Dirac quantization, physical states are annihilated by all first-class constraints. Second-class
constraints cannot be imposed this way as they do not commute amongst themselves.
The physical states in the presence of second-class constraints can be found by imposing
the generalized Gupta-Bleuler conditions,\cite{Allen:1988hc,Kalau:1991jp,Bellucci:1997dr} 
which are that the physical matrix elements of all second-class constraints vanish. 
Thus, the constraint matrix elements between physical states must satisfy
\begin{eqnarray*}
%\label{eq:PhiElements}
\langle \phi | \hat{\varphi}_1 | \psi \rangle &=&  - (\phi_3^*\psi_1 -
\phi_1^*\psi_3) - \frac{ 1}{ 2\mathstrut} (\phi_0^*\psi_2 - \phi_2^*\psi_0) = 0  \> , \cr 
\langle \phi | \hat{\varphi}_2 | \psi \rangle &=& - (\phi_3^*\psi_2 -
\phi_2^*\psi_3) + \frac{1}{2} (\phi_0^*\psi_1 - \phi_1^*\psi_0) =0  \>   . 
\end{eqnarray*}
Because the matrix elements of a Grassmann-odd operator between two states of the
same Grassmann parity vanish automatically, the task of identifying physical states
reduces to finding states of definite Grassmann parity such that matrix elements of the
constraints between states of opposite parity vanish. We find that
\begin{eqnarray}
  \langle \phi_{\rm even} | \hat\varphi_1 | \psi_{\rm odd} \rangle &=&
  - (\phi_3^*\psi_1) - \frac{1}{2\mathstrut} (\phi_0^*\psi_2) = 0 \>  , \cr
  \langle \phi_{\rm even} | \hat\varphi_2 | \psi_{\rm odd} \rangle &=&
  - (\phi_3^*\psi_2) + \frac{1}{2} (\phi_0^*\psi_1) = 0  \>  ,
\end{eqnarray}
are satisfied when $2 (\phi_3/\phi_0)^* = - (\psi_2/\psi_1) =
(\psi_1/\psi_2)$. 

These conditions yield an orthonormal basis for the wave
functions of the full Schr\"odinger state space,
\begin{eqnarray}\label{eq:fullbasis}
|{0}\rangle &=&1 + \frac{i}{2}\xi_1\xi_2 \>  , \cr
|{1}\rangle&=&\frac{1}{\sqrt2}(\xi_1 + i\xi_2) \>  , \cr
|\bar{0}\rangle&=&1 - \frac{i}{2}\xi_1\xi_2  \> , \cr
|\bar{1}\rangle&=&\frac{1}{\sqrt2}(\xi_1 - i\xi_2)  \>  .
\end{eqnarray}
The inner product (\ref{eq:SchrodNorm}) gives
\begin{equation}
\langle \alpha|\beta \rangle = \delta_{\alpha\beta} = - \langle \bar{\alpha} | \bar{\beta} \rangle \> ,
%  \langle 0 | 0 \rangle = \langle 1 | 1 \rangle = 1 = - \langle \bar{0} |
% \bar{0} \rangle = - \langle \bar{1} | \bar{1} \rangle ,
\end{equation}
for $\alpha=0,1$ and $\beta=0,1$, and the matrix elements of the constraints
vanish in the physical basis $|\alpha\rangle$,
\begin{equation}
\langle \alpha | \hat\varphi_k | \beta \rangle = 0 \> ,
\end{equation}
decomposing the full Schr\"od\-inger Hilbert space into a physical space 
and an orthogonal negative-norm ``ghost'' space,
\begin{equation}\label{eq:SchroHilbert}
  {\mathfrak H}_{\text{Schr\"odinger}} = {\mathfrak H}_{\rm physical} \oplus {\mathfrak
    H}_{\rm ghost}  \> .
\end{equation}
The constraint operators each map the physical state space to the ghost
state space and vice versa:
\begin{eqnarray}
%  \hat\varphi_1 | 0 \rangle &=& - \frac{i}{\sqrt2} |\bar{1} \rangle ,\qquad
%  \hat\varphi_1 | 1 \rangle = + \frac{i}{\sqrt2} | \bar 0 \rangle , \cr 
%  \hat\varphi_2 | 0 \rangle &=& + \frac{1}{\sqrt2} |\bar{1} \rangle ,\qquad
%  \hat\varphi_2 | 1 \rangle = - \frac{1}{\sqrt2} | \bar 0 \rangle , \cr 
%  \hat\varphi_1 | \bar 0 \rangle &=& - \frac{i}{\sqrt2} |{1} \rangle ,\qquad
%  \hat\varphi_1 | \bar 1 \rangle = + \frac{i}{\sqrt2} | 0 \rangle , \cr 
%  \hat\varphi_2 | \bar 0 \rangle &=& - \frac{1}{\sqrt2} |{1} \rangle ,\qquad
%  \hat\varphi_2 | \bar 1 \rangle = + \frac{1}{\sqrt2} | 0 \rangle .\cr
 \hat\varphi_k | \alpha \rangle &=& - \frac{i^k}{\sqrt2} \epsilon_{\alpha\beta} | \bar{\beta} \rangle \>  , \cr
   \hat\varphi_k |\bar{\alpha} \rangle  &=& \frac{(- i)^k}{\sqrt2} \epsilon_{\alpha\beta} |{\beta} \rangle \>  ,
\end{eqnarray}
where $\epsilon_{01} = -\epsilon_{10} = 1$ and $\epsilon_{00} = \epsilon_{11} = 0$.

The remaining condition on an inner product is the self-adjointness of all observables. 
Anticommuting variables, being nilpotent, cannot be observables, 
but nonetheless $\hat{\xi}_i$ is self-adjoint,
\begin{equation}
  \left(\xi_i\psi(\xi_1,\xi_2)\right)^* = \psi(\xi_1,\xi_2)^* \xi_i \> ,
\end{equation}
and we have $\langle \phi | \hat{\xi}_i \psi \rangle = \langle \hat{\xi}_i
\phi | \psi \rangle $, or $\hat{\xi}_i^\dagger = \hat{\xi}_i$. 
Although $\xi_i$ is real, its conjugate momentum $\pi_i$ is purely imaginary.
The momentum operator $\hat{\pi}_i$ should therefore be anti-self-adjoint, 
which one can check by direct calculation:
\begin{equation}
  \int \phi^* \left(i{\partial^L\over\partial\xi_i} \psi\right) d\xi_1\, d\xi_2
  = - \int  \left(i{\partial^L\over\partial\xi_i}\phi\right)^*\psi
  \> d\xi_1\, d\xi_2  \> ,
\end{equation}
and so $\hat{\pi}_i^\dagger = - \hat{\pi}_i $. 

Up to a constant factor, the only observable in this system is the Hamiltonian corresponding to
Eq.~(\ref{eq:Hthree}),
\begin{equation}\label{eq:HtwoOp}
  \hat{H}^\prime =  i\omega\left(\hat\pi_1 + \frac{i}{2}\hat\xi_1\right)\left(\hat\pi_2 + \frac{i}{2}\hat\xi_2\right)  \> ,
\end{equation}
which is manifestly self-adjoint.

Because the Hamiltonian (\ref{eq:HtwoOp}) comes from the Grassmann-even
Hamiltonian (\ref{eq:Hthree}), the physical eigenstates can be taken to have
definite Grassmann parity. We find
\begin{eqnarray}\label{eq:GB2spectrum}
  \hat{H}^\prime | 0 \rangle &=&  -\frac{\omega}{2\mathstrut} | 0 \rangle \>  , \cr
  \hat{H}^\prime | 1 \rangle &=&  \frac{\omega}{2} | 1 \rangle \>  .
\end{eqnarray}
The ghost states, though unphysical, are also eigenstates of the Hamiltonian (\ref{eq:HtwoOp}).

\subsection{Comparison with Reduced Phase Space Quantization}

\subsubsection{Dirac brackets and the reduced phase space}

Since the physical evolution of a classical constrained system must remain on the
``constraint surface'' where the second-class constraints vanish, it is possible
to set the constraints identically to zero both inside and outside of Poisson 
brackets and work purely with functions on the constraint surface, the 
reduced phase space. We examine this approach to quantization here. 
Consistency requires that the Poisson bracket on the full 
phase space be replaced by the Dirac bracket \cite{Dirac:1964:LQM} on the 
reduced phase space, 
\begin{equation}\label{eq:genDB}
  \{f,g\}_{DB} = \{f,g\} - \{f,\varphi_n\}\Delta^{nm}\{\varphi_m,g\} \> ,
\end{equation}
where $\Delta^{nm}$ is the inverse matrix to $\{\varphi_n,\varphi_m\}$. The
constraints $\varphi_m \approx 0$ can be taken to be strongly zero
because the Dirac bracket of anything with a constraint vanishes
identically,
\begin{eqnarray}
  \{f,\varphi_k\}_{DB} \strut &=& \{f,\varphi_k\} -
  \{f,\varphi_n\}\Delta^{nm}\{\varphi_m,\varphi_k\} \cr
  &=& \{f,\varphi_k\} -
  \{f,\varphi_n\}\delta^{n}_{k} \equiv 0 \>  .
\end{eqnarray}
The Dirac bracket has the same symmetry properties as the Poisson bracket
and satisfies the Jacobi identity. 

In our case, the matrix of Poisson brackets of the constraints is
\begin{equation}
  \{\varphi_k,\varphi_\ell\} = -i\delta_{k\ell} \> , 
\end{equation}
so the Dirac bracket becomes
\begin{equation}
  \{f,g\}_{DB} = \{f,g\} - i \{f,\varphi_k\} \{\varphi_k,g\} \>  .
\end{equation}
The full phase space is four-dimensional while the constraint surface is
two-dimensional. The two coordinates $\xi_1$ and $\xi_2$ can be used
as phase space coordinates on the constraint surface. Their Dirac brackets
are 
\begin{eqnarray}
  \{\xi_i,\xi_j\}_{DB} &=& \{\xi_i ,\xi_j \} - i \{\xi_i,\varphi_k\}   \{\varphi_k, \xi_j\} \cr 
  &=& 0^{\strut} - i \delta_{ik} \delta_{kj} = - i \delta_{ij} \> ,
\end{eqnarray}
so that the Dirac bracket of functions $f(\xi_1,\xi_2)$ and
$g(\xi_1,\xi_2)$ on the constraint surface is
\begin{equation}\label{eq:RPSDBone}
 \{f,g\}_{DB} = -i\sum_{k=1,2}\left({\partial^R f\over \partial \xi_k}{\partial^L
   g\over \partial \xi_k}\right) \> .
\end{equation}

% We note that up to an overall sign, the Dirac bracket
% Eq.~(\ref{eq:RPSDBone}) is the abstract Poisson bracket postulated by
% Berezin and Marinov, at least for two anticommuting variables.

\subsubsection{Operators and states}
As $\xi_1$ and $\xi_2$ are coordinates of the two-dimensional reduced phase
space, for quantization in the Schr\"odinger picture one must choose one
position coordinate and one canonical momentum to proceed. Neither $\xi_1$ nor
$\xi_2$ can fulfill either role as each has non-vanishing Dirac bracket with
itself.

Instead, following the holomorphic representation,\cite{Berezin:1961ct,Faddeev:1980be} we consider the complex phase space coordinates,
\begin{eqnarray}\label{eq:EtaBarEta}
  \eta &=& \frac{1}{\strut\sqrt2}(\xi_1 + i \xi_2)  \> , \cr
  \bar\eta &=& \frac{1}{\sqrt2}(\xi_1 - i \xi_2) \>  , 
\end{eqnarray}
which satisfy
\begin{equation}
  \{\eta,\eta \}_{DB} = \{\bar\eta,\bar\eta \}_{DB} = 0 \> , \quad
  \{\eta,\bar\eta \}_{DB} = - i \> .
\end{equation}
For quantization we need operators that satisfy the Dirac anticommutation relations,
\begin{equation}
 \hat{\bar\eta}\hat\eta+\hat{\eta}\hat{\bar\eta}  = i\hbar\widehat{ \{\bar\eta,\eta \}}_{DB} =  \hbar \>  ,
\end{equation}
and can proceed with quantization in the Schr\"odinger picture by taking
states to be functions of $\eta$ alone,
\begin{equation}
  \psi = \psi(\eta) = \psi_0 + \psi_1\eta  \> ,
\end{equation}
and the operators $\hat\eta$ and
$\hat{\bar\eta}$ acting upon them to be
\begin{eqnarray}
      \hat\eta &=& \eta \>  , \cr
\hat{\bar\eta} &=& \hbar {\partial^L\over \partial \eta} \> .
\end{eqnarray}
Again, we use units in which $\hbar=1$ in what follows.  As we are working
in the reduced phase space, the constraints were eliminated before
quantization, so all we need do now is construct the inner product and find
the spectrum.

\subsubsection{Inner product}

While the wave function is a function of $\eta$, its complex conjugate is a function of
$\bar\eta$. Thus it is necessary to consider inner products of holomorphic form,\cite{Berezin:1961ct,Faddeev:1980be}
\begin{equation}\label{eq:RPSinner}
  \langle \phi | \psi \rangle = \int \phi^*(\bar\eta) \psi(\eta) {\mathcal
    M}(\bar\eta,\eta)\, d\eta\,d\bar\eta \>  ,
\end{equation}
where ${\mathcal M}(\bar\eta,\eta)$ is a measure factor needed to enforce the
adjointness relations coming from the complex conjugate nature of the
variables $\bar\eta$ and $\eta$, $\eta^* = \bar\eta$. We need to have
\begin{equation}
  \hat\eta^\dagger = \hat{\bar\eta} = {\partial^L\over \partial \eta} \> ,
\end{equation}
or, for any two states $\phi$ and $\psi$,
\begin{eqnarray}\label{eq:etaadjoint}
  \langle \hat{\bar\eta} \phi | \psi \rangle &=& \int \left({\partial^L \phi\over
    \partial \eta}\right)^* \psi(\eta)\, {\mathcal M}(\bar\eta,\eta)\, d\eta\,d\bar\eta \cr
  & = &   \strut\int \phi^*(\bar\eta)\, \eta\psi(\eta)\, {\mathcal M}(\bar\eta,\eta)\,
  d\eta\,d\bar\eta\, \cr 
  & = & \strut\langle \phi | \hat\eta \psi \rangle \> .
\end{eqnarray}
Similarly, it is necessary that $\hat\eta$ = $\hat{\bar\eta}^\dagger$, or
\begin{eqnarray}\label{eq:etabaradjoint}
  \langle \hat{\eta} \phi | \psi \rangle 
  &=& \int ( \eta\phi)^* \psi(\eta)\, {\mathcal M}(\bar\eta,\eta)\, d\eta\,d\bar\eta \cr
  &=& \int \phi^*(\bar\eta)\bar\eta\, \psi(\eta)\, {\mathcal M}(\bar\eta,\eta)\, d\eta\,d\bar\eta \cr
  &=&\int\phi^*(\bar\eta)\,{\partial^L\psi\over\partial\eta}\, {\mathcal M}(\bar\eta,\eta)\,
  d\eta\,d\bar\eta\, \cr 
  & = & \langle \phi | \hat{\bar\eta} \psi \rangle \> .
\end{eqnarray}
%Putting the general ${\mathcal M}(\bar\eta,\eta)={\mathcal M}_{00}+ {\mathcal
%  M}_{10}\bar\eta+ {\mathcal M}_{01}\eta+ {\mathcal M}_{11}\bar\eta \eta $ into the
%conditions, we find that t
For the adjointness conditions Eqs.~(\ref{eq:etaadjoint}) and (\ref{eq:etabaradjoint})
to hold, it is necessary and sufficient that up to an overall factor, 
\begin{equation}\label{eq:RPSmeasure}
  {\mathcal M}(\bar\eta, \eta) = 1 + \bar\eta\eta = \exp(\bar\eta\eta) \> .
\end{equation}
With the measure Eq.~(\ref{eq:RPSmeasure}), the inner product on states $\psi(\eta)=\psi_0 + \psi_1\eta$ and
$\phi(\eta) = \phi_0 + \phi_1\eta$ is
\begin{equation}
  \langle \phi | \psi \rangle = \int \phi^* \psi \, \exp(\bar\eta\eta)\,
  d\eta\,d\bar\eta = \psi_0^*\phi_0 + \psi_1^* \phi_1 \> .
\end{equation}
This inner product leads to positive definite norms for states.
Because the constraints have been implemented prior to quantization,
there is no ghost sector.

Note that the basis eigenstates in this system, $|0\rangle$ and
$|1\rangle$, have wave functions, $1$ and $\eta$ respectively, with 
definite Grassmann parity that correspond to the Grassmann parities 
of the equivalent states found under Gupta-Bleuler quantization. 
The similarity between the states of the two different quantizations 
is stronger than just their Grassmann parities, however.

\subsubsection{Gupta-Bleuler and Reduced Phase Space Wave Function Correspondence}

The Gupta-Bleuler configuration space, rather than the reduced phase space, can also be
parametrized by the $\eta$ and $\bar\eta$ coordinates of
Eq.~(\ref{eq:EtaBarEta}), allowing us to rewrite the physical Gupta-Bleuler
wave functions given in Eq.~(\ref{eq:fullbasis}) as the reduced phase space
ones times the square root of the reduced phase space measure factor,
\begin{eqnarray}\label{eq:GBRPSmap}
|0\rangle &=& 1 + \frac{i}{2\strut}\xi_1\xi_2 = 1 + \frac{1}{2}\bar\eta\eta 
=  \sqrt{e^{\bar\eta\eta}} = \sqrt{{\mathcal M}} \> , \cr
|1\rangle &=& \frac{1}{\sqrt2}(\xi_1 + i\xi_2) = \eta
 = \eta \sqrt{e^{\bar\eta\eta}} =  \eta\sqrt{{\mathcal M}} \> ,
\end{eqnarray}
which is to say
\begin{equation}
  \label{eq:1}
 \left(\strut\psi_n(\xi_1,\xi_2)\right)_{\rm GB} = \left(\psi_n(\eta) \sqrt{{\mathcal M}}\right)_{\rm RPS} \>  .
\end{equation}
The inner product on the physical Gupta-Bleuler space of states is
the integral over the $\xi_1$, $\xi_2$ configuration space, which can be
reparametrized as an integral over the $\eta$, $\bar\eta$ reduced phase
space, making the orthonormality of the one set understandable in terms of
the other.

\subsection{The Primed Variables of Hanson, Regge \& Teitelboim}

Instead of replacing the Poisson brackets with Dirac brackets in a system with second-class constraints,
Hanson, Regge, and Teitelboim \cite{Hanson:1976cn} show that one can replace the canonical variables,
or indeed any dynamical variables, by primed versions that agree on the constraint surface 
and whose Poisson brackets with any other quantity also agree with the Dirac brackets of 
those quantities on the constraint surface. These are the so-called primed variables,
\begin{equation}
A^\prime = A - \{A,\varphi_n\}\Delta^{nm}\varphi_m \approx A \> ,
\end{equation}
where again $\Delta^{nm}$ is the inverse matrix to $\{\varphi_n,\varphi_m\}$. The Dirac
bracket satisfies the weak equalities \cite{Hanson:1976cn} 
\begin{equation}
\{A,B\}_{DB} \approx  \{A^\prime,B^\prime\} \approx  \{A^\prime,B\} \approx \{A,B^\prime\} \> .
\end{equation}
Because the matrix of Poisson brackets of the constraints is constant, the 
primed versions of the $\xi_i$ variables with the constraints (\ref{eq:constraintsone}), 
\begin{equation}\label{eq:HRTxiprime}
\xi_k^\prime = \frac12 \xi_k - i\pi_k \>  ,
\end{equation}
have strongly vanishing Poisson brackets with the constraints
\begin{equation}
\{\varphi_j,\xi^\prime_k\} = 0 \> .
\end{equation}
These primed variables will be important for extending this system to general Hamiltonians.

\section{$D=2N$ Anticommuting Variables}\label{sec:evenDs}

To handle the $D=2N$ case, we generalize
the measure for computing inner products of wave functions analogously to the construction for commuting 
coordinates,
\begin{equation}\label{eq:evenDmeasure}
d\mu_{2N} = d\mu_{2N-2} \left(i\, d\xi_{2N-1} d\xi_{2N}\right) = i^N \, d\xi_1 \ldots d\xi_{2N} \> .
\end{equation}

As we already have the quantization for $D=2$ case, we here show how to go from $D=2N-2$ to $D=2N$. Suppose that 
$\Psi^+_{2N-2}(\xi_1,\xi_2, \ldots \xi_{2N-2})$ is a positive norm physical state for $D=2N-2$ variables with definite Grassmann parity. Then if $\psi^+_{2N}(\xi_{2N-1},\xi_{2N})$ is also a positive norm physical state of definite Grassmann parity for a system consisting of just the two variables $\xi_{2N-1}$ and $\xi_{2N}$, then 
\begin{equation}
|++\,\rangle_{2N} = \Psi^+_{2N-2}\psi^+_{2N}
\end{equation}
is a positive norm physical state of definite Grassmann parity for the system consisting of $D=2N$ anticommuting variables. 
 
 Similarly, suppose that $\Psi^-_{2N-2}(\xi_1,\xi_2, \ldots \xi_{2N-2})$ is a negative norm ghost state for $D=2N-2$ variables with definite Grassmann parity. Then if $\psi^-_{2N}(\xi_{2N-1},\xi_{2N})$ is a negative norm ghost state of definite Grassmann parity for a system consisting of just the two variables $\xi_{2N-1}$ and $\xi_{2N}$, then 
\begin{equation}
|--\,\rangle_{2N} = \Psi^-_{2N-2}\psi^-_{2N}
\end{equation}
 is also positive norm physical state of definite Grassmann parity for the system consisting of $D=2N$ anticommuting variables. One has to check that the norms work as stated and that the Gupta-Bleuler conditions hold. This is straightforward, if tedious.  Similarly, the states
 \begin{eqnarray}
|+- \, \rangle_{2N} &=& \Psi^+_{2N-2}\psi^-_{2N} \>  , \cr
|-+ \, \rangle_{2N} &=& \Psi^-_{2N-2}\psi^+_{2N} \>   
 \end{eqnarray}
are negative norm ghost states. Thus we have $2^{2N-1} = 2^{D-1}$ physical states and an equal number of ghost states. The total number of physical and ghost states is $2^{D-1} + 2^{D-1}=2^D$, the total number of terms in a function of $D$ anticommuting variables.

\section{ $D=2N+1$ Anticommuting Variables}\label{sec:oddDs}

\subsection{One anticommuting variable, $N=0$}\label{sec:trivial}

We begin by considering the special case of a single real anticommuting variable,
with an eye to the general case. The case of a single anticommuting variable has also 
been examined by Bordi and Casalbuoni,\cite{Bordi:1980aj} and Bordi, 
Casalbuoni, and Barducci.\cite{Barducci:1980rx}

With only one anticommuting variable, and absent anticommuting constant parameters, the only possible term in the Lagrangian is the
kinetic term,
\begin{equation}\label{eqn:L1}
  L = \frac{i}{2} \xi\dot\xi  \> .
\end{equation}
The equation of motion for $\xi$ is that it is a constant.
The momentum of the system does not depend on the velocity, 
\begin{equation}
  \pi = {\partial^R L \over \partial \dot\xi} = \frac{i}{2}\xi \> ,
\end{equation}
so there is a constraint
\begin{equation}\label{eq:oneDconstraint}
  \varphi = \pi - \frac{i}{2}\xi \approx 0 \> ,
\end{equation}
and the only dynamics are that the system obeys the constraint, because the
Hamiltonian vanishes identically. The phase space consists of the single
variable $\xi$. Effectively there is just ``half a degree of freedom.''
The only way to quantize this system is to use the Gupta-Bleuler
quantization to impose the constraint. In the Schr\"odinger representation,
the wave function is a linear function of $\xi$,
\begin{equation}
  \psi(\xi) = \psi_0 + \psi_1\xi \> ,
\end{equation}
with $\psi_0$ and $\psi_1$ being complex numbers. Because the Poisson
brackets, as we will see, are $\{\pi,\xi\} = \{\xi,\pi\} = 1$, the Dirac
quantization rule gives the momentum operator (in $\hbar = 1$ units)
\begin{equation}
\hat\pi = i{\partial^L\over\partial\xi} \> .  
\end{equation}

In analogy to the quantum mechanics of one commuting variable, we set the inner product to be the integral
\begin{equation}\label{eq:oneDnorm}
  \langle \phi | \psi \rangle = \int \phi^*(\xi) \psi(\xi) \, d\xi =
  \phi_1^*\psi_0 + \phi_0^*\psi_1 \> .
\end{equation}
Since the variable $\xi$ is real,
$(\phi_0 + \phi_1\xi)^* = \phi_0^* + \phi_1^*\xi$.  Gupta-Bleuler
quantization requires the constraint to have vanishing matrix elements
between any two physical states, namely
\begin{equation}\label{eq:phiEV}
  \langle\phi | \hat\varphi | \psi\rangle = i(\phi_1^* \psi_1 -
  \frac{1}{2}\phi_0^*\psi_0) = 0 \> ,
\end{equation}
which implies that up to an overall phase, there is just a single
normalized physical state of positive norm,
\begin{equation}\label{eq:halfDoFstate}
  \psi_{\rm phys}(\xi) = \frac{1}{\sqrt[4]{2}}\left(1 + \frac{1}{\sqrt{2}}\xi\right) \> ,
\end{equation}
and a single orthogonal ghost state of negative norm,
\begin{equation}\label{eq:halfDoFghost}
  \psi_{\rm ghost}(\xi) = \frac{1}{\sqrt[4]{2}}\left(1 - \frac{1}{\sqrt{2}}\xi\right) \> .
\end{equation}
The states are eigenstates of $\sqrt2\hat\xi^\prime$ with eigenvalues $\pm 1$.
\begin{eqnarray}
\sqrt2\hat\xi^\prime \psi_{\rm phys}(\xi) &=& + \psi_{\rm phys}(\xi) \> , \cr
\sqrt2\hat\xi^\prime \psi_{\rm ghost}(\xi) &=& - \psi_{\rm ghost}(\xi) \> .
\end{eqnarray}

Note that if we were to take our single anticommuting variable to be
imaginary, $\zeta = i\xi$, we would change the sign of the kinetic term. This 
will be important in considering the Lorentzian case.  

It is  important to 
mention that keeping the Lagrangian as (\ref{eqn:L1}) with a positive 
overall sign but positing an imaginary $\xi$, or equivalently, having a 
negative sign for the Lagrangian and a real $\xi$, makes it impossible 
to impose the constraint (\ref{eq:oneDconstraint}) through the integral 
(\ref{eq:phiEV}) because that expression becomes proportional to a 
positive definite expression, $\langle \psi |\hat\varphi | \psi \rangle \propto \psi_1^*\psi_1 +
\psi_0^*\psi_0/2$.

%Note, finally, that we can convert this to a model with a single imaginary
%anticommuting variable by defining $\xi^\prime=i\xi$.  This changes the
%overall sign of the Lagrangian, and will lead to some factors of $i$ in the
%inner product and the physical state and will change the relative sign in
%the constraint.  This will play a role in Sections \ref{sec:lorentzian} and
%\ref{sec:fourlorentzian}. It is also worth mentioning that keeping the
%Lagrangian as (\ref{eqn:L1}) with a positive overall sign but positing an
%imaginary $\xi$, or having a negative sign for the Lagrangian and a real
%$\xi$, makes it impossible to impose the constraint
%(\ref{eq:oneDconstraint}) through the integral (\ref{eq:phiEV}) because
%that expression becomes proportional to a positive definite expression,
%$\langle \psi |\hat\varphi | \psi \rangle \propto \psi_1^*\psi_1 +
%\psi_0^*\psi_0/2$.

\subsection{Three anticommuting variables}

We now generalize to the case of three anticommuting variables, the first
case treated by Berezin and Marinov,\cite{Berezin:1976eg} handled here in our Schr\"odinger formalism. After
diagonalization of the kinetic terms, the most general Lagrangian is
\begin{equation}
  L = \frac{i}{2} \xi_k \dot\xi_k + i \omega_k \epsilon_{ijk} \xi_i \xi_j \>  ,
\end{equation}
which contains three arbitrary commuting constants, $\omega_k$.  A further
rotation of the $\xi_i$ and $\omega_k$ allows the
reduction of the
Lagrangian to
\begin{equation}\label{eq:LthreeBthree}
  L = \frac{i}{2} \xi_k \dot\xi_k + i \omega \xi_1 \xi_2 \>  ,
\end{equation}
which has the same form as the Lagrangian (\ref{eq:Lagrangian}), except now
the kinetic term contains the additional piece
$\frac{i}{2} \xi_3 \dot\xi_3$.  As a consequence, we might try to
anticipate the result of the explicit quantization.  Since the Lagrangian
(\ref{eq:LthreeBthree}) separates into two non-interacting parts, one
involving $\xi_1$ and $\xi_2$ and having the form of the two-variable
system analyzed earlier, and the other involving $\xi_3$
and having the form analyzed in the preceding section, the basis states
of the three-variable system can be written in terms of products of the
basis states of those two simpler systems.  The Hamiltonian that commutes
with the constraints will be identical to (\ref{eq:Hthree}).

%In particular, we might expect that the basis states for the physical
%sector of the three-variable system can be obtained by taking a product of
%two physical basis states or two ghost basis states, one from the
%two-variable system and one from the one-variable system.  Likewise, the
%ghost basis states for the three-variable system would be written as the
%product of a physical basis state and a ghost basis state, one from the
%two-variable system and one from the one-variable system.  It turns out
%that this is a correct description of what happens with the
%Gupta-Bleuler quantization, but there is a reduced phase space
%approach that gives a slightly different result, as we will see.

Note that when we compare the three-variable system to the two-variable
system, two of the constraints and two of the equations of motion are the
same but there is one additional constraint, which has the same form as the
other two constraints,
\begin{equation}
  \varphi_3 = \pi_3 - \frac{i}{2}\xi_3 \approx 0 \> ,
\end{equation}
and one additional equation of motion,
\begin{equation}
\dot\xi_3 = 0 \> .  
\end{equation}
As we know, the one-variable system has a Hamiltonian that
vanishes identically, and so the Hamiltonian for the three-variable
system has the same form as Eq.~(\ref{eq:Hthree}), although
the wave functions can have $\xi_3$ dependence.

We now give the results of explicit quantization.

\subsubsection{States}

With three Grassmann coordinates, the ``measure'' will now be the 
Grassmann-odd product $i\,d\xi_1\,d\xi_2\,d\xi_3$. With this 
measure, normalizable states cannot have a definite 
Grassmann parity. For the system described by (\ref{eq:LthreeBthree}), 
the wave functions of the system can be factorized as
\begin{equation}\label{eq:threexiPsi}
  \Psi(\xi_1,\xi_2,\xi_3) = \psi(\xi_1,\xi_2)\,u(\xi_3)  \> .
\end{equation}
If the two-dimensional wave functions $\psi(\xi_1,\xi_2)$ have definite
Grassmann parity, then it is easy to see that the matrix elements of the
first two second-class constraints will vanish if the two-dimensional
factors of the wave functions are either both in
$\mathfrak{H}_{\rm physical}$ or both in $\mathfrak{H}_{\rm ghost}$ of the
two-variable Hilbert space (\ref{eq:SchroHilbert});
\begin{eqnarray}
  \langle \Phi | \hat\varphi_{1,2} | \Psi \rangle &=& i \int
  (\phi\, v)^* \hat\varphi_{1,2} \psi\, u
  \, d\xi_1\,d\xi_2\,d\xi_3 \cr 
  & = &  i \int (v^* \tilde{u}) (\phi^* \hat\varphi_{1,2} \psi)\,
  \, d\xi_1\,d\xi_2\,d\xi_3\cr
  & = & 0 \> ,
\end{eqnarray}
where $\tilde u$ is either $u(-\xi_3)$ or $u(\xi_3)$, depending on whether
the Grassmann parities of $\phi(\xi_1,\xi_2)$ and $\psi(\xi_1,\xi_2)$ are
the same or different respectively.
The matrix elements of
the third constraint are
\begin{eqnarray}\label{eq:PhiThree}
  \langle \Phi | \hat\varphi_{3} | \Psi \rangle &=& i \int
  (\phi\,v)^* \hat\varphi_{3} \psi\, u
  \, d\xi_1\,d\xi_2\,d\xi_3 \cr 
&=& i \int  v^*\,\phi^* \hat\varphi_{3} \psi\, u
  \, d\xi_1\,d\xi_2\,d\xi_3 \cr 
  & = &  i \int  (v^* \hat\varphi_3 \tilde u)\, (\phi^*\psi)
  \, d\xi_1\,d\xi_2\,d\xi_3\cr
  & = & i \int (v^* \hat\varphi_3 \tilde u)\, d\xi_3  \int  \phi^*\psi\,
  \, d\xi_1\,d\xi_2\, ,\quad\strut
\end{eqnarray}
where $\tilde u(\xi_3)$ is $(-1)^{g_\phi}u((-1)^{g_\phi+g_\psi}\xi_3)$,
where $g_\phi$ and $g_\psi$ denote the Grassmann parities of $\phi$ and
$\psi$, respectively. The second factor,
$\int \phi^*\psi\, d\xi_1\,d\xi_2$, vanishes unless $g_\phi=g_\psi$. In
that case, $\tilde u = (-1)^{g_\phi}u$ and there are two solutions that
make the matrix elements (\ref{eq:PhiThree}) vanish, both of which have
$u=v$, namely
\begin{equation}\label{eq:xi3physical}
  u(\xi_3) = v(\xi_3) = \frac{1}{\sqrt[4]{2}}\left(1 \pm \frac{\xi_3}{\sqrt2}\right) \>  .
\end{equation}

The norm of a product wave function $\psi(\xi_1,\xi_2)\, u(\xi_3)$ is the
product of the norms of its factors,
\begin{eqnarray}
  \langle \psi\, u | \psi\, u \rangle & = & i \int u^*\, \psi^* \psi\, u  \,
  d\xi_1\,d\xi_2\,d\xi_3 \cr
  &=& \left(\int u^*
  u\, d\xi_3\right) \left(i \int \psi^* \psi \, d\xi_1\,d\xi_2\right)  \> . \qquad\strut
\end{eqnarray}
Consequently, the positive norm physical states are spanned by the orthonormal basis
\begin{eqnarray}\label{eq:threexiphys}
 |0\rangle & = & \frac{1}{\sqrt[4]2}\left(1 + \frac{i}{2}\xi_1\xi_2\right)\left (1 +
 \frac{\xi_3}{\sqrt2}\right) \>  ,\cr  
 |1\rangle & = & \frac{1}{\sqrt[4]8} \left(\xi_1 + i\xi_2\right)\left(1 + \frac{\xi_3}{\sqrt2}\right) \> ,\cr
 |0^\prime\rangle & = & \frac{1}{\sqrt[4]2}\left(1 - \frac{i}{2}\xi_1\xi_2\right) \left(1 -
\frac{\xi_3}{\sqrt2}\right) \> , \cr 
|1^\prime\rangle & = & \frac1{\sqrt[4]8} \left(\xi_1 - i\xi_2\right)\left(1 - \frac{\xi_3}{\sqrt2}\right) \> .
\end{eqnarray}

The negative norm ghost states are spanned by the orthogonal anti-normal basis
\begin{eqnarray}
 |\bar{0}\rangle  & = & \frac{1}{\sqrt[4]2}\left(1 - \frac{i}{2}\xi_1\xi_2\right) \left(1 +
 \frac{\xi_3}{\sqrt2}\right) \> , \cr  
 |\bar{1}\rangle  & = & \frac1{\sqrt[4]8} \left(\xi_1 - i\xi_2\right)\left(1 + \frac{\xi_3}{\sqrt2}\right) \> , \cr
 |\bar{0}^\prime\rangle & = & \frac{1}{\sqrt[4]2}\left(1 + \frac{i}{2}\xi_1\xi_2\right)
 \left(1 - \frac{\xi_3}{\sqrt2}\right) \> , \cr 
 |\bar{1}^\prime\rangle & = & \frac1{\sqrt[4]8} \left(\xi_1 + i\xi_2\right)\left(1 - \frac{\xi_3}{\sqrt2}\right) \>  . 
\end{eqnarray}

The large Schr\"odinger Hilbert space once again splits as in
Eq.~(\ref{eq:SchroHilbert}), but this time both the physical and ghost
Hilbert spaces have dimension four. Each of these Hilbert spaces is the reducible
$\bf{2}\oplus\bf{2}$ representation of the three-dimensional Clifford
algebra. 

\subsubsection{Physical spectrum}

The states (\ref{eq:threexiphys}) are eigenstates of the Hamiltonian:
%(\ref{eq:Hthree}).
\begin{eqnarray}\label{eq:threexiphyseigen}
 \hat{H}_{\rm phys}|0\rangle & =  -\frac{\textstyle\strut\omega}{\strut\textstyle 2} |0\rangle  \> ,\quad 
   \hat{H}_{\rm phys}|0^\prime\rangle & =  -\frac{\textstyle\omega}{\textstyle 2} |0^\prime\rangle \>  ,\cr
 \hat{H}_{\rm phys}|1\rangle & =  +\frac{\textstyle\strut\omega}{\strut\textstyle 2} |1\rangle \>  , \quad 
   \hat{H}_{\rm phys}|1^\prime\rangle & =  +\frac{\textstyle\strut\omega}{\textstyle 2} |1^\prime\rangle \>  .
\end{eqnarray}

\subsubsection{Matrix elements of $\hat{\xi}_i$ and Pauli algebra}

We find the matrix elements of the position operators $\hat{\xi}_i$ in the physical basis $\{|0\rangle, |1\rangle \}$ to be
%\begin{eqnarray}
 % \pmatrix{ \langle 0 | \hat\xi_1 | 0 \rangle & \langle 0 | \hat\xi_1 | 1 \rangle
 %   \cr
 %   \langle 1 | \hat\xi_1 | 0 \rangle & \langle 1 | \hat\xi_1 | 1 \rangle
 %   \cr}  &=& \frac{1}{\sqrt2\strut} \pmatrix{ 0 & 1 \cr 1 & 0\cr} , \cr
 % \pmatrix{ \langle 0 | \hat\xi_2 | 0 \rangle & \langle 0 | \hat\xi_2 | 1 \rangle
 %   \cr
 %   \langle 1 | \hat\xi_2 | 0 \rangle & \langle 1 | \hat\xi_2 | 1 \rangle
 %   \cr}  &=& \frac{1}{\sqrt2\strut} \pmatrix{ 0 & i \cr - i & 0\cr} ,\cr
 %  \pmatrix{ \langle 0 | \hat\xi_3 | 0 \rangle & \langle 0 | \hat\xi_3 | 1 \rangle
 %   \cr
 %   \langle 1 | \hat\xi_3 | 0 \rangle & \langle 1 | \hat\xi_3 | 1 \rangle
 %   \cr}  & = & \frac{1\strut}{\sqrt2\strut} \pmatrix{ 1 & 0 \cr 0 & - 1 \cr} .
%\end{eqnarray}
\begin{equation}\label{eq:PauliEVs} 
\langle \alpha | \strut \, \hat \xi_k| \beta \rangle = \frac{1\strut}{\sqrt2\strut}\, (\sigma_k)_{\beta\alpha} \> ,
\end{equation}
where $\sigma_k$ are the standard Pauli matrices. 
It is instructive to note that the diagonal entries in $\sigma_3$ in 
Eq.~(\ref{eq:PauliEVs}) result from the even or odd definite
Grassmann parities of the $\psi(\xi_1,\xi_2)$ pieces (\ref{eq:threexiPsi})
of the basis states $|0\rangle$ and $|1\rangle$ of
(\ref{eq:threexiphys}). 

While the matrix elements of the $\hat\xi_k$ yield the Pauli matrices, 
the $\hat\xi_k$ operators themselves do not form a Clifford algebra; 
they are still nilpotent generators of a Grassmann algebra. However, 
the $\hat\xi_k^\prime$ operators corresponding to Eq.~(\ref{eq:HRTxiprime})
do form a Clifford algebra, {although one not obeying a definite Pauli 
algebra, either $\sqrt2\hat\xi_i^\prime\hat\xi_j^\prime = + {i} \epsilon_{ijk}\hat\xi_k^\prime$ or $\sqrt2\hat\xi_i^\prime\hat\xi_j^\prime = - {i} \epsilon_{ijk}\hat\xi_k^\prime$, unless
one restricts to a superselection sector. In the physical sectors that Pauli algebra is left-handed, while in
the ghost sectors it is right-handed.}

\subsection{General $D=2N+1$}

{As with even dimensions, for the general odd dimensional case, we generalize the measure for computing inner products of wave functions analogously to the construction for commuting coordinates, setting}
\begin{equation}\label{eq:oddDmeasure}
d\mu_{2N+1} = d\mu_{2N} \left(d\xi_{2N+1}\right) = i^N \, d\xi_1 d\xi_2 \ldots d\xi_{2N+1} \> .
\end{equation}

As we already have the quantization for $D=2N$ case, we show here how to go from $D=2N$ to $D=2N+1$. Suppose that 
$\Psi^+_{2N}(\xi_1,\xi_2, \ldots \xi_{2N})$ is a positive norm physical state for $D=2N$ variables with definite Grassmann parity. Then if $\psi^+_{2N+1}(\xi_{2N+1})$ is  the positive norm physical state (which, as we saw, must have mixed Grassmann parity) for a system consisting of just the single variables $\xi_{2N+1}$, then 
\begin{equation}
|++\,\rangle_{2N+1} = \Psi^+_{2N}\psi^+_{2N+1}
\end{equation}
is a positive norm physical state of mixed Grassmann parity for the system consisting of $D=2N+1$ anticommuting variables. 
 
Similarly, suppose that $\Psi^-_{2N}(\xi_1,\xi_2, \ldots \xi_{2N})$ is a negative norm ghost state for $D=2N$ variables with definite Grassmann parity. When $\psi^-_{2N}(\xi_{2N-1},\xi_{2N})$ is the negative norm ghost state for a system consisting of just the variable $\xi_{2N+1}$, then 
\begin{equation}
|--\,\rangle_{2N+1} = \Psi^-_{2N}\psi^-_{2N+1}
\end{equation}
 is also a positive norm physical state of mixed Grassmann parity for the system consisting of $D=2N+1$ anticommuting variables. Again, one has to check that the norms work as stated and that the Gupta-Bleuler conditions hold, which is somewhat tedious.  Similarly, the states
 \begin{eqnarray}
|+- \, \rangle_{2N+1} &=& \mathstrut\Psi^+_{2N}\psi^-_{2N+1} \>  , \cr
|-+ \, \rangle_{2N+1} &=& \Psi^-_{2N}\psi^+_{2N+1} \>   
 \end{eqnarray}
are negative norm ghost states. Thus we have $2^{2N} = 2^{D-1}$ physical states and an equal number of ghost states. The total number of physical and ghost states is $2^{D-1} + 2^{D-1}=2^D$, the total number of terms in a function of $D$ anticommuting variables. 

\section{Lorentzian metrics}\label{sec:lorentzian}

If the metric for the $\xi$ variables is not Euclidean but Lorentzian with 
signature $(-,+,+,+,\ldots)$, we can map the system in variables $\xi_0,\xi_1,\ldots\xi_{D-1}$ 
to the Euclidean case with variables $\xi_1,\xi_2,\ldots\xi_{D}$, for instance 
by defining a new real Grassmann variable $\xi_D =  i\xi_0$. We saw at 
the end of section \ref{sec:trivial} that a time-like Grassmann variable must be 
imaginary, or no physical states of just that one variable can exist.  {With this 
redefinition to map to the Euclidean case, the analysis can then proceed for the}
$D$ real Grassmann variables $\xi_1,\xi_2,\ldots\xi_{D}$ with Euclidean signature.
In $D=3+1$ dimensions, for example, in terms of the original Lorentzian variables 
the positive norm physical states are {
\begin{eqnarray} \label{eq:LorentzFourBasis}
  |0\rangle &=&\strut\left(1+\frac{i}{\strut2}\xi_1\xi_2\right)\left(1  + \frac{1}{\strut2}\xi_0\xi_3\right) \> ,\cr
  |1\rangle &=&\frac{1}{\strut\sqrt2}\left(\xi_1+i\xi_2\right)\left(1 {+ }\frac{1}{2}\xi_0\xi_3\right)  \> ,\cr
  |2\rangle &=&\frac{1}{\strut\sqrt2}\left(1+\frac{i}{2}\xi_1\xi_2\right) \left(\xi_3 -\xi_0\right) \> ,\cr
  |3\rangle &=&\frac{1}{2\strut}\left(\xi_1+i\xi_2\right)\left(\xi_3 -\xi_0\right) \> ,\cr
  |0^\prime\rangle &=&\left(1-\frac{i}{2}\xi_1\xi_2\right)\left(1 - \frac{1}{2}\xi_0\xi_3\right) \> ,\cr
  |1^\prime\rangle &=&\frac{1}{\strut\sqrt2}\left(\xi_1 - i\xi_2\right)\left(1 - \frac{1}{2}\xi_0\xi_3\right)  \> ,\cr
  |2^\prime\rangle &=&\frac{1}{\strut\sqrt2}\left(1-\frac{i}{2}\xi_1\xi_2\right) \left(\xi_3 +\xi_0\right) \> ,\cr
  |3^\prime\rangle &=&\frac{1}{2}\left(\xi_1 - i\xi_2\right)\left(\xi_3  + \xi_0\right) \> ,
 \end{eqnarray}
while the ghost states are
\begin{eqnarray}
  |\bar 0\rangle &=&\left(1+\frac{i}{2}\xi_1\xi_2\right)\left(1  - \frac{1}{2}\xi_0\xi_3\right) \> ,\cr
  |\bar 1\rangle &=&\frac{1}{\strut\sqrt2}\left(\xi_1+i\xi_2\right)\left(1 {- }\frac{1}{2}\xi_0\xi_3\right)  \> ,\cr
  |\bar 2\rangle &=&\frac{1}{\strut\sqrt2}\left(1 +\frac{i}{2}\xi_1\xi_2\right) \left(\xi_3 +\xi_0\right) \> ,\cr
  |\bar 3\rangle &=&\frac{1}{\strut2}\left(\xi_1+i\xi_2\right)\left(\xi_3 +\xi_0\right) \> ,\cr
  |\bar 0^\prime\rangle &=&\left(1-\frac{i}{2}\xi_1\xi_2\right)\left(1 + \frac{1}{2}\xi_0\xi_3\right) \> ,\cr
  |\bar 1^\prime\rangle &=&\frac{1}{\strut\sqrt2}\left(\xi_1 - i\xi_2\right)\left(1 + \frac{1}{2}\xi_0\xi_3\right)  \> ,\cr
  |\bar 2^\prime\rangle &=&\frac{1}{\strut\sqrt2}\left(1-\frac{i}{2}\xi_1\xi_2\right) \left(\xi_3 -\xi_0\right) \> ,\cr
  |\bar 3^\prime\rangle &=&\frac{1}{2}\left(\xi_1 - i\xi_2\right)\left(\xi_3  - \xi_0\right) \> .
\end{eqnarray}
}

\section{Superselection sectors}

\subsection{Even dimensions}

As explained in Sec.~(\ref{sec:evenDs}), in the case that $D$ is even,  the physical wave functions are the products
\begin{equation}\label{eq:physgenD}
\psi_{\rm phys}(\xi_1,\xi_2,\ldots\xi_{D}) =   \prod_{{n=1}\atop{\prod s_m = 1}}^{D/2} \psi^{s_n}_{i_n}(\xi_{2n-1},\xi_{2n})
\end{equation}
%\begin{eqnarray}
 % \psi_{i_1}^+(\xi_1,\xi_2)  \psi_{i_2}^+(\xi_3,\xi_4) \psi_{i_3}^+(\xi_5,\xi_6) \cdots \psi_{i_{[N/2]}}^+(\xi_{N-1},\xi_N) , \nonumber \\
 % \psi_{i_1}^{-}(\xi_1,\xi_2)  \psi_{i_2}^{-}(\xi_3,\xi_4) \psi_{i_3}^+(\xi_5,\xi_6) \cdots \psi_{i_{[N/2]}}^+(\xi_{N-1},\xi_N) ,\nonumber \\
 % \psi_{i_1}^{-}(\xi_1,\xi_2)  \psi_{i_2}^{-}(\xi_3,\xi_4) \psi_{i_3}^-(\xi_5,\xi_6) \cdots \psi_{i_{[N/2]}}^+(\xi_{N-1},\xi_N) , \nonumber 
%\end{eqnarray}
%etc.,
{of 2D wave functions with an even number of ghost factors. Here the superscript $s_n$ is the sign of the norm of the state, with $s_n=+1$ for physical states, and $s_n=-1$ for ghost states. Because ghost states are orthogonal to physical states, we can see that the Hilbert space of physical states is a sum of superselection sectors}
\begin{equation}
  {\mathfrak H}_{\rm phys} = \hskip -20pt \bigoplus^{\lfloor D/2\rfloor}_{k = 0 \atop 1\leq j_1 < j_2 < \cdots <j_{2k}\leq\lfloor D/2\rfloor } \hskip - 20pt {\mathfrak H}_{ j_1 \, j_2 \, \cdots \, j_{2k}} \> ,
\end{equation}
where ${\mathfrak H}_{ i_1 \, i_2 \, \cdots \, i_{2k}}$ is spanned by products  of the form (\ref{eq:physgenD}) with the negative norm factors in ``slots'' $j_1$, $j_2$, $\ldots$ , $j_{2k}$. In other words, $s_{j_1} = s_{j_2} = \ldots  = s_{j_{2k}} = -1$ and the rest of the $s_m = +1$. Each of the superselection sectors ${\mathfrak H}_{ i_1 \, i_2 \, \cdots \, i_{2k}}$  has dimension $2^{\lfloor D/2 \rfloor}$,
\begin{equation}
\dim {\mathfrak H}_{ i_1 \, i_2 \, \cdots \, i_{2k}} = 2^{\lfloor D/2 \rfloor} \> ,
\end{equation}
 and there are
\begin{equation}
\sum_{n=0}^{2\lfloor D/4\rfloor} \left( \lfloor D/2\rfloor \atop 2n \right) = 2^{\lfloor D/2\rfloor - 1 }
\end{equation}
different superselection sectors.

The ghost states are of similar form,
\begin{equation}
\psi_{\rm ghost}(\xi_1,\xi_2,\ldots\xi_{D}) =   \prod_{n=1\atop \prod s_m = -1}^{D/2} \psi^{s_n}_{i_n}(\xi_{2n-1},\xi_{2n}) \> ,
\end{equation}
but with an odd number of negative norm factors. There are also $2^{\lfloor D/2\rfloor - 1 }$ different ghost superselection sectors.

\subsection{Odd dimensions}

In the case of odd $D$, we introduce the obvious notation for the single-variable wave functions $\psi^s(\zeta)$ of Eqs.~(\ref{eq:halfDoFstate}) and (\ref{eq:halfDoFghost}),
\begin{equation}
  \psi^\pm(\zeta) = \frac{1}{\sqrt[4]{2}}\left( 1 \pm \frac{1}{\sqrt2}\zeta\right) \>  ,
\end{equation}
that have inner products
\begin{equation}
  \int \psi^{s_1 *}(\zeta) \psi^{s_2}(\zeta) \, d\zeta = s_1 \delta_{{s_1}{s_2}} \>  .
\end{equation}
The physical states can be seen as states of form 
\begin{equation}
\psi_{\rm phys}(\xi_1,\ldots,\xi_{D}) = \left(\prod_{n=1\atop \prod s_m = \strut\, s_D}^{\lfloor D/2\rfloor} \psi^{s_n}_{i_n}(\xi_{2n-1},\xi_{2n})\right) \psi^{s_D}(\xi_{D}) \>  ,
\end{equation}
{while the ghost states satisfy the product condition} $\prod s_m = -\, s_D$.
There are now $2^{\lfloor D/2\rfloor}$ physical superselection sectors, {twice as many as in the even $D$ case.} Each superselection sector has
dimension $2^{\lfloor D/2\rfloor}$ as {in the even $D$ case. The counting is the same for the ghost states.}

\section{General Hamiltonians and stability of the Gupta-Bleuler conditions}

One can check that the $\hat{\xi}_i^\prime$ operators acting on a physical
state yield a physical state, and acting on a ghost state yield a ghost
state; that is, the $\hat{\xi}_i^\prime$ operators do not change the sign of the
norm of a state. Further, they map any superselection sector, whether ghost or physical, onto itself. 
In contrast, the constraint operators $\hat{\varphi}_i$
do change the sign of the norm of a state, so they map physical states into
ghost states, and vice versa. The $\hat{\xi}_i^\prime$ and
$\hat{\varphi}_j$ operators anticommute with each other,
\begin{equation}
  \hat{\xi}_i^\prime \hat{\varphi}_j +  \hat{\varphi}_j \hat{\xi}_i^\prime = 0  \>  ,
\end{equation}
which can be expressed through the commutative diagram \ref{diagram:xiphi} below.
\begingroup
\def\normalbaselines{\baselineskip20pt \lineskip3pt \lineskiplimit3pt }
\def\mapright#1{\smash{ \mathop{\longrightarrow }\limits^{#1}}}
\def\mapleft#1{\smash{ \mathop{\longleftarrow }\limits^{#1}}}
\def\mapdown#1{\Big\downarrow \rlap{$\vcenter{\hbox{$\scriptstyle #1$}}$}}
\def\mapup#1{\Big\uparrow \rlap{$\vcenter{\hbox{$\scriptstyle #1$}}$}}
\begin{equation}\label{diagram:xiphi}
\begin{matrix}
{\mathfrak H}_{\rm phys} &\mapright{\hat{\xi}_i^\prime\strut}&{\mathfrak H}_{\rm phys}\\
        \mapdown{\hat \varphi_j}&&\mapdown{\hat\varphi_j}\\
        {\mathfrak H}_{\rm ghost}&\mapright{-\hat\xi_i^\prime\strut}&{\mathfrak H}_{\rm ghost}\\
        \mapdown{\hat \varphi_j}&&\mapdown{\hat\varphi_j}\\
        {\mathfrak H}_{\rm phys} &\mapright{\hat\xi_i^\prime\strut}&{\mathfrak H}_{\rm phys}
\end{matrix}
\end{equation}
\endgroup
The maps are all onto and, if done twice, give back the state scaled by 1/2, as expressed by the anticommutation relations
\begin{equation}\label{eq:cliffordxiprime}
 \hat\xi_i^\prime\hat\xi_j^\prime + \hat\xi_j^\prime\hat\xi_i^\prime = \delta_{ij}  \> 
\end{equation}
and
\begin{equation}
 \hat\varphi_i\hat\varphi_j+ \hat\varphi_j\hat\varphi_i = \delta_{ij}  \> ,
\end{equation}
which extend the commutative diagram \ref{diagram:xiphi} to the larger toroidal diagram \ref{diagram:xiphi2} below.
\begin{center}
\begin{equation}\label{diagram:xiphi2}
\begin{tikzpicture}[node distance=2cm, auto]
% \node (P){ ${\mathfrak H}_{\rm phys}$};
  \node (A) {${\mathfrak H}_{\rm phys}$};
  \node(B) [right of=A]  {${\mathfrak H}_{\rm phys}$};
  \node(C) [right of=B]  {${\mathfrak H}_{\rm phys}$};
  \node (D) [below of=A]{${\mathfrak H}_{\rm ghost}$};
  \node (E) [below of=B] {${\mathfrak H}_{\rm ghost}$};
  \node (F) [below of=C] {${\mathfrak H}_{\rm ghost}$};
  \node (G) [below of=D]{${\mathfrak H}_{\rm phys}$};
  \node (H) [below of=E] {${\mathfrak H}_{\rm phys}$};
  \node (I) [below of=F] {${\mathfrak H}_{\rm phys}$};
  \draw[->] (A) to node [swap] {${\hat{\xi}^\prime_i}$} (B);
  \draw[->] (B) to node [swap] {${\hat\xi^\prime_i}$} (C);
  \draw[->] (D) to node {$-{\hat\xi^\prime_i}$} (E);
  \draw[->] (E) to node {$-{\hat\xi^\prime_i}$}(F);
  \draw[->] (G) to node {${\hat\xi^\prime_i}$}(H);
  \draw[->] (H) to node {${\hat\xi^\prime_i}$}(I);
  \draw[->] (A) to node  {${\hat\varphi_j}$} (D);
  \draw[->] (B) to node  {${\hat\varphi_j}$} (E);
  \draw[->] (C) to node  [swap]{${\hat\varphi_j}$} (F);
  \draw[->] (D) to node  {${\hat\varphi_j}$} (G);
  \draw[->] (E) to node  {${\hat\varphi_j}$} (H);
  \draw[->] (F) to node  [swap]{${\hat\varphi_j}$} (I);
  \draw[->, bend right] (A) to node [swap] {$\frac{1}{2}\cdot \textbf{id}$} (G);
  \draw[->, bend left] (C) to node  {$\frac{1}{2} \cdot \textbf{id}$} (I);
  \draw[->, bend left] (A) to node {$\frac{1}{2} \cdot \textbf{id}$} (C);
  \draw[->, bend right] (G) to node [swap] {$\frac{1}{2}\cdot \textbf{id}$} (I);
\end{tikzpicture}
\end{equation}
\end{center}

In either commutative diagram \ref{diagram:xiphi} or \ref{diagram:xiphi2}, the space ${\mathfrak H}_{\rm phys}$ can be any
physical superselection sector ${\mathfrak H}_{ i_1 \, i_2 \, \cdots \, i_{2k}}$ or any sum of physical superselection sectors, 
so it is consistent to restrict the physical space to any of the isomorphic physical superselection sectors,  ${\mathfrak H}_{ i_1 \, i_2 \, \cdots \, i_{2k}}$, yielding the physical Hilbert space as a
$2^{\lfloor D/2\rfloor }$-dimensional irreducible module of the Clifford algebra generated by the $\hat\xi_i^\prime$,  Eq.~(\ref{eq:cliffordxiprime}).

Because the general Hamiltonian $\hat{H}_{\rm phys} = H(\hat{\xi}^\prime)$, as in the $D=2$ case Eq.~(\ref{eq:Hthree}), is built
from the primed $\hat{\xi}_i^\prime$ operators, the energy eigenstates will span
the physical Hilbert space, no matter which of the superselection sectors is chosen for the physical Hilbert space. The Gupta-Bleuler conditions,
\begin{equation}\label{eq:GBgeneral}
  \langle \phi_{\rm phys} | \hat{\varphi}_i | \psi_{\rm phys}\rangle = 0  \>  ,
\end{equation}
will thus hold for all times if they hold at any one time, because the time
evolution of an initially physical state in one superselection sector remains in the same physical superselection sector. Thus, the matrix
elements (\ref{eq:GBgeneral}) are zero for all time.

The commutative diagram \ref{diagram:xiphi2} also makes clear that there is a symmetry between the physical and ghost sectors. The determination of which sectors are physical and which are ghost is an artifact of the choice of the sign of the inner product. If the other sign of the inner product (from Eq.~(\ref{eq:evenDmeasure}) or Eq.~(\ref{eq:oddDmeasure})) is chosen, the physical and ghost sectors are swapped.

\section{Equivalence to Dirac-K\"ahler Fermions}\label{sec:Kahler}

Wave functions taking values in the space of antisymmetric tensors, equivalent to being valued in the 
space of differential forms, and a wave equation for a spin 1/2 particle in terms of them has a very long 
history, going back to Ivanenko and Landau\cite{Ivanenko:1928zf} in 1928. In the early 
1960s, K\"ahler \cite{Kahler:1960zz,Kahler:1961as,Kahler:1962id} found a mapping of the Dirac equation onto
inhomogeneous differential forms. These ideas have been further developed by Graf \cite{Graf:1978kr} and many others. 
Dirac-K\"ahler fermions have also been proposed \cite{Becher:1982ud} as a solution to the fermion doubling problem on the lattice. 
The existence of the identical sectors was noticed by Benn and Tucker\cite{Benn:1982sr};
Banks, Dothan, and Horn \cite{Banks:1982iq}; and Becher and Joos.\cite{Becher:1982ud}
In $D=3+1$, the four sectors---in our case, two physical and two ghost---have been posited \cite{Banks:1982iq} 
as a solution to the family problem of the standard model.  Jourjine\cite{Jourjine:2019jxn} 
argued recently that when the masses of the four generations are degenerate or sufficiently close, 
only three of the generations are observable and the fourth is hidden. 

The space of complex-valued functions of $D$ Grassmann variables, ${\mathscr F}_{D}$,
%\begin{equation}
%{\mathscr F}_{D} = \big\{ f(\xi_1,\ldots,\xi_D) = {\textstyle \sum }_{k=0}^D a_{\mu_1\mu_2\cdots\mu_k}\xi^{\mu_1}\xi^{\mu_2}\cdots\xi^{\mu_k} \mid a_{\mu_1\mu_2\cdots\mu_k} \in {\mathbb C}  \big\}, 
%\end{equation}
is isomorphic to the space of inhomogeneous
 differential forms on $\mathbb{R}^D$ with complex coeffcients,
\begin{equation}
{\mathscr F}_D \cong \Lambda_{\mathbb{C}}(\mathbb{R}^D) = \bigoplus_{p=0}^D \Lambda^p_{\mathbb{C}}(\mathbb{R}^D) \> ,
\end{equation}
by the bijection
\begin{eqnarray}
&& a_{\mu_1\mu_2\cdots\mu_k}\xi^{\mu_1}\xi^{\mu_2}\cdots\xi^{\mu_k} \longleftrightarrow \cr
&&  a_{\mu_1\mu_2\cdots\mu_k}({\sqrt2})^k\, dx^{\mu_1}\wedge dx^{\mu_2} \wedge \cdots\wedge dx^{\mu_k}  \>  .
\end{eqnarray}
%\begin{eqnarray}
%\xi^\mu &\longleftrightarrow & {\sqrt2}\, dx^\mu, \cr
%\sum_{k=0}^D \frac{1}{k!} a_{\mu_1\cdots \mu_k} \xi^{\mu_1}\cdots\xi^{\mu_k} &\longleftrightarrow& \sum_{k=0}^D \frac{1}{k!} a_{\mu_1\cdots %\mu_k} \left({\sqrt2}\right)^k dx^{\mu_1}\wedge\cdots \wedge dx^{\mu_k},
%\end{eqnarray}
%where the $k=0$ term is simply a constant and the $k=1$ term has but one index. 

Left multiplication of a function on Grassmann space by a Grassmann variable $\xi^\nu$, which is the effect of the operator $\hat\xi^\nu$, corresponds to wedging the corresponding form by $dx^\nu$,
\begin{equation}
\hat\xi^\nu \longleftrightarrow {\sqrt2}\, dx^\nu\wedge  \>  ,
\end{equation}
while the action of the associated momentum, $\hat\pi_\nu = i {\partial^L\over\partial\xi^\nu}$, corresponds to a contraction with the corresponding vector,
\begin{equation}
\hat\pi_\nu \longleftrightarrow i \frac{1}{\sqrt2} e_\nu\inner \> ,
\end{equation}
where $e_\nu\inner dx^\mu = \delta^\mu_\nu$.  Thus the operator for the primed coordinates, $\hat\xi^{\prime\mu} = {\partial^L\over\partial \xi_\mu} + \frac{1}{2}\eta^{\mu\nu}\xi_\nu $, corresponds exactly to the Clifford product\cite{Kahler:1960zz,Kahler:1961as,Kahler:1962id} applied to the corresponding form,
\begin{equation}
\hat\xi^{\prime}\mathstrut^{\mu} \longleftrightarrow \frac{1}{\sqrt2}\, dx^\mu\vee \> .
\end{equation}
Specifically, under this correspondence we have
\begin{equation}
\varphi\left(\hat\xi^{\prime}\mathstrut^{\mu}\psi\right) =  \frac{1}{\sqrt2}\, dx^\mu\vee \varphi(\psi) \> ,
\end{equation}
and
\begin{equation}
\varphi\left(\sqrt2\hat\xi^{\prime}\mathstrut^{\mu}\partial_\mu\psi\right) =   dx^\mu\vee \partial_\mu\varphi(\psi) \>.
\end{equation}

When the commuting variables $x^\mu$ are added to the action\cite{Berezin:1976eg}, the wave functions become functions of both the $x^\mu$ and the $\xi_\nu$. The inner product between states becomes
\begin{equation}
\langle \phi | \psi \rangle = \int_{\mathbb{R}^D} d^Dx\,\, i^{\lfloor {D\over 2}\rfloor}\int \phi^*(x,\xi) \,\psi(x,\xi) \,  d\xi_1\cdots d\xi_D \> .
\end{equation}
If we denote the action of the bijection $\varphi \colon {\mathscr F}_D \to \Lambda_{\mathbb{C}}(\mathbb{R}^D)$ on an element $\psi \in {\mathscr F}_D$ as $\varphi(\psi)$, the inner products in the two spaces, which lead to indefinite norms, are related by 
\begin{equation}\label{eq:innerDKtranslation}
\langle \phi | \psi \rangle = {{i^{\lfloor \frac{D}{2}\rfloor}}{2^{-{D}/{2}}}}\int_{\mathbb{R}^D} \varphi(\phi^*) \wedge\varphi(\psi)  \> .
\end{equation}

This inner product differs from the usual positive definite inner product for Dirac-K\"ahler wave functions given by
\begin{equation}\label{eq:innerDK} 
\langle \varphi(\phi)| \varphi(\psi) \rangle = \int_{\mathbb{R}^D} \overline{\varphi(\phi)} \wedge\ast\,\varphi(\psi)  \> , 
\end{equation}
where $\overline{\varphi(\phi)}$ is the complex conjugate of the form ${\varphi(\phi)}$  and $\,\ast\, $ is the Hodge star operator that maps $p$-forms to $(D-p)$-forms.   Note that $\varphi(\phi^*) = (-1)^{p(p-1)/2}\overline{\varphi(\phi)} $ when $\varphi(\phi^*)$ is a form of degree $p$. Manko\v{c} Bor\v{s}tnik and Nielsen\cite{MankocBorstnik:1999th} have also found the relationship between Grassmann quantum mechanical wave functions and Dirac-K\"{a}hler fermions, but do so from the operatorial point of view and take the standard positive-definite inner product, (\ref{eq:innerDK}).

\section{Discussion and Conclusions}

Quantized pseudoclassical systems in the Schr\"odinger realization using
the generalized Gupta-Bleuler method exhibit rich interdependences among the
reality of the variables, the Grassmann parity of the wave functions, and
the split between physical and ghost states, though this structure has heretofore been hidden because
the quantization of pseudoclassical theories in the Schr\"odinger realization  has
been relatively less studied in comparison to the path integral quantization. The
existence of the Schr\"odinger realization has been assumed by Bordi,
Casalbuoni, and Barducci,\cite{Bordi:1980aj,Barducci:1980rx} who also
first found the physical states given in Eq.~(\ref{eq:halfDoFstate}), 
and by Manko\v{c} Bor\v{s}tnik.\cite{MankocBorstnik:1992np} 
Delbourgo \cite{Delbourgo:1987ye} considered
nonrelativistic spin systems represented by the Schr\"odinger picture
quantum mechanics of two anticommuting variables, and relativistic systems
represented by four anticommuting variables. He also considered more
general involutions on these variables.

The present quantization of pseudoclassical theories in the Schr\"odinger realization using Dirac's machinery for constrained systems appears to
be the first to examine these systems not purely in terms of operators and their representations, but also to construct explicit wave functions
and an explicit indefinite inner product, and to examine the adjointness properties of the operators under that inner product following from the involution
properties of the Grassmann variables. The present quantization also explicitly realizes the Dirac-K\"ahler formulation of fermions in the language of Grassmann 
calculus rather than differential forms; the two descriptions are isomorphic.  

In $D=3+1$, the physical states (\ref{eq:LorentzFourBasis}) have been looked 
at from a more abstract operatorial point of view by Manko\v{c} Bor\v{s}tnik
\cite{MankocBorstnik:1993ia,MankocBorstnik:1994hh} and Manko\v{c}
Bor\v{s}tnik and Nielsen,
\cite{MankocBorstnik:1999th,MankocBorstnik:2001tn,MankocBorstnik:2003fr}
which is closer to our Gupta-Bleuler quantization than to the
reduced phase space quantization that the abstract approach of Berezin and
Marinov \cite{Berezin:1976eg} most closely resembles.  Crucially, Manko\v{c}
Bor\v{s}tnik and Nielsen\cite{MankocBorstnik:1999th} examine operators, 
$\tilde{a}^a$ and ${\tilde{\tilde a}}\mathstrut^a$, that play a central role in their analysis and 
correspond (up to constant factors) to our constraints $\hat\varphi_j\approx 0$
and primed variables $\hat\xi^\prime_j$, respectively.

We have seen that in this Gupta-Bleuler quantization, adding one more
real Grassmann coordinate to a system with an even number of Grassmann
variables has two effects. The first is that the number of physical states
will double because the ghost state for the new variable can pair with
ghost states of the previous system to make physical states in the combined
system. In terms of the quantum mechanics, these new states are in a
different superselection sector; one can then choose whether to include one or both
of these superselection sectors. The second effect is to
make the physical states be of mixed Grassmann parity, because the
``measure'' in the integral defining the inner product will now have odd Grassmann parity.  By
contrast, the reduced phase space quantization has a positive definite
inner product and so always produces an irreducible representation of the
Clifford algebra; adding one more Grassmann coordinate to the system does
not lead to a doubling of the number of physical states and there are no superselection sectors
in a reduced phase space quantization. This should not be surprising as it is 
well known\cite{Romano:1989zb,Schleich:1990gd,Allen:1992sp} that 
reduced phase space quantizations are not always equivalent to other 
quantizations of the same constrained system.

We have also seen that the behavior of the Grassmann coordinates under the
involution, in other words, whether the variables are taken to be real or imaginary,
has an effect on the quantum system. In the trivial case, the quantum
mechanics of an imaginary Grassmann variable cannot have a Schr\"odinger
realization unless the kinetic term is negative because the constraint
cannot otherwise be imposed. 

Since the behavior of the pseudoclassical variables under the involution 
determines the adjointness properties of
the corresponding quantum operators, the timelike $\xi_0$ must have
reality properties opposite to the spacelike $\xi_i$, and their
corresponding quantum operators, the gamma matrices $\gamma^0$ and
$\gamma^i$, therefore must have opposite self-adjointness properties. 

We have not paid much attention in the present work to systems with more than just anticommuting degrees of freedom, nor have we considered the physical meaning of 
the Lorentzian case, but Berezin and Marinov\cite{Berezin:1976eg} argue 
that while the $\xi_0$ needs to be present for manifest Lorentz invariance, 
one needs to remove it dynamically by imposing a pseudoclassical ``Dirac 
equation''  constraint, which dynamically removes the $\xi_0$ from the 
system in a covariant way.  Doing so leads to the Dirac equation and world-line supersymmetry. In the present work we have not examined the 
rich systems with both anticommuting and commuting variables, only mentioning 
them in passing in Sec.~\ref{sec:Kahler} in order to make a connection to Dirac-K\"ahler fermions.
Nonetheless, our methods can be fruitfully brought to bear on the full
range of actions that include both commuting and anticommuting classical variables.

\vfil
\appendix
\section{$D=3+1$ Gamma matrix representation}

In the $3+1$ dimensional case, the primed variables lead to operators
\begin{eqnarray}
 \xi^{\prime}\mathstrut^{\mu} &=& \xi^\mu -
 \{\xi^\mu,\varphi_\alpha\}\Delta^{\alpha\beta}\varphi_\beta = \xi^\mu
                     -i(\pi^\mu - \frac{i}{2}\xi^\mu ) \>  , \cr
 \hat\xi^{\prime}\mathstrut^{\mu} &=& {\partial^L\over\partial \xi_\mu} +
                         \frac{1}{2}\eta^{\mu\nu}\xi_\nu  \>  , 
\end{eqnarray}
that satisfy the
anticommutation relations  
\begin{equation}
\hat\xi^{\prime}\mathstrut^{\mu}\hat\xi^{\prime}\mathstrut^{\nu} +
\hat\xi^{\prime}\mathstrut^{\nu}\hat\xi^{\prime}\mathstrut^{\mu} = \eta^{\mu\nu} = {\rm
  diag}(-,+,+,+)  \>  .
\end{equation}
The $\hat\xi^{\prime}\mathstrut^{\mu}$ can be
represented by scaled Dirac gamma matrices; $\sqrt2\,\hat\xi^{\prime}\mathstrut^{\mu} \to
\gamma^\mu$.
In the unprimed (\textit {i.e.\/} $|\kern -2pt ++\rangle$ sector) physical basis (\ref{eq:LorentzFourBasis}), we define
\begin{eqnarray}
\psi_0|0\rangle + \psi_1|1\rangle + \psi_2|2\rangle + \psi_3|3\rangle =
 \left(
 \begin{matrix}
  \psi_0 \\ \psi_1\\ \psi_2\\ \psi_3\\  
  \end{matrix}
\right)
\end{eqnarray}
and find that the representation of
the $\hat\xi^{\prime\mu}$ in this basis is {
\begin{eqnarray}  \label{eq:gammarep}
  \sqrt2\,\hat\xi^{\prime}\mathstrut^0 & = &\left(\>\begin{matrix} 0 & \phantom{-}0 &                      -1 &  \phantom{-}0 \\ 
                                                                                                                              0 & \phantom{-}0 & \phantom{-}0 & \phantom{-}1 \\
                                                                                                                              1 & \phantom{-}0 & \phantom{-}0 & \phantom{-}0 \\ 
                                                                                                                              0 &                      -1 & \phantom{-}0 & \phantom{-}0 \end{matrix}\>\right)     
                                                                                                                        = - i{\sigma_2}\otimes {\sigma_3}  \> , \cr 
  \sqrt2\,\hat\xi^{\prime}\mathstrut^1 & = &\left(\>\begin{matrix} 0 & \phantom{-}1 & \phantom{-}0 & \phantom{-}0 \\
                                                                                                                              1 & \phantom{-}0 & \phantom{-}0 & \phantom{-}0 \\ 
                                                                                                                              0 & \phantom{-}0 & \phantom{-}0 & \phantom{-}1 \\ 
                                                                                                                              0 & \phantom{-}0 & \phantom{-}1 & \phantom{-}0 \end{matrix}\>\right)  
                                                                                                                       = \mathds{1}\otimes\sigma_1   \> ,\cr 
  \sqrt2\,\hat\xi^{\prime}\mathstrut^{2} & = & \left(\begin{matrix} 0 & \phantom{-}i & \phantom{-}0 & \phantom{-}0   \\ 
                                                                                                                             -i & \phantom{-}0 &\phantom{-}0 & \phantom{-}0  \\ 
                                                                                                                             0 & \phantom{-}0  &\phantom{-}0  &  \phantom{-}i   \\ 
                                                                                                                             0 & 0 & - i & \phantom{-}0 \end{matrix}\>\right) 
                                                                                                                       = -\mathds{1}\otimes \sigma_2 \> , \cr 
  \sqrt2\,\hat\xi^{\prime}\mathstrut^{3} & = & \left(\>\begin{matrix}  0 & \phantom{-}0 & \phantom{-}1 & \phantom{-}0  \\ 
                                                                                                                                   0 & \phantom{-}0 & \phantom{-}0 &                       -1 \\ 
                                                                                                                                   1 & \phantom{-}0 & \phantom{-}0 &  \phantom{-}0 \\ 
                                                                                                                                   0 &             -1          & \phantom{-}0  & \phantom{-}0 \end{matrix}\>\right)
                                                                                                                        = {\sigma_1}\otimes {\sigma_3}  \> .
\end{eqnarray}
}
\begin{acknowledgments}
  T.J.A. thanks J. R. Schmidt for useful discussions and library
  assistance. C.W. acknowledges support from the Provost's Office 
  at Hobart and William Smith Colleges and the Carey-Cohen Fund 
  for Summer Research.
\end{acknowledgments}

\end{document}